\documentclass[10pt,english,aps,showkeys,showpacs,nofootinbib]{revtex4-1}
\usepackage[T1]{fontenc}
\usepackage{float}
\usepackage{textcomp}
\usepackage{amsmath}
\usepackage{amssymb}
\usepackage{graphicx}
\makeatletter

\usepackage[utf8]{inputenc}
\usepackage{todonotes}
\usepackage[bottom]{footmisc}
\usepackage{graphicx}
\usepackage{float}
\usepackage{dcolumn}
\usepackage{bm}
\usepackage{todonotes}
\usepackage{hyperref}
\usepackage[scale=0.75]{geometry}
\@ifundefined{showcaptionsetup}{}{%
 \PassOptionsToPackage{caption=false}{subfig}}
\usepackage{subfig}
\makeatother
\usepackage{babel}
\usepackage{rotating}
\begin{document}
\title{Dynamics of a Klein-Gordon Oscillator (KGO) in the Presence of a Cosmic
String in Som\textendash Raychaudhuri Space-Time}
\author{Abdelmalek Bouzenada }
\email{abdelmalek.bouzenada@univ-tebessa.dz ; abdelmalekbouzenada@gmail.com}

\affiliation{Laboratory of theoretical and applied Physics,~\\
 Echahid Cheikh Larbi Tebessi University, Algeria}
\author{Abdelmalek Boumali}
\email{boumali.abdelmalek@gmail.com}

\affiliation{Laboratory of theoretical and applied Physics,~\\
 Echahid Cheikh Larbi Tebessi University, Algeria}
\author{R. L. L .Vit\'{o}ria}
\email{ricardo.vitoria@pq.cnpq.br ; ricardo-luis91@hotmail.com}

\affiliation{Faculdade de F\'{\i}sica, Universidade Federal do Par\'{a}, Av. Augusto
Corr\^{e}a,~\\
 Guam\'{a}, Bel\'{e}m, PA 66075-110, Brazil}
\author{C. Furtado}
\email{furtado@fisica.ufpb.br}

\affiliation{Departamento de F\'{\i}sica, Universidade Federal da Para\'{\i}ba,~\\
 Caixa Postal 5008, Jo\~{a}o Pessoa-PB 58051-900, Brazil}

\date{\today}
\begin{abstract}
This paper explores the dynamics of the Klein-Gordon oscillator in the presence of a cosmic string in Som Raychaudhuri spacetime. The exact solutions for the free case and the oscillator case are obtained and discussed. These solutions reveal the effects of the cosmic string and the spacetime geometry on the bosonic particles. To illustrate these results, some figures and tables have been included.
\end{abstract}
\maketitle

\section{Introduction }

An in-depth investigation into the impact of the gravitational field on the dynamics of quantum mechanical systems remains a topic of great interest. Albert Einstein's general theory of relativity (GR) \citep{key-1} offers an engaging and all-encompassing portrayal of gravity as an intrinsic geometric feature of spacetime. This theory sheds light on the profound link between the curvature of spacetime and the emergence of the classical gravitational field, resulting in highly successful predictions of remarkable phenomena such as gravitational waves \citep{key-2} and black holes \citep{key-3}. Concurrently, quantum mechanics (QM) provides a robust framework for comprehending and describing the intricate behavior of particles at the microscopic level \citep{key-4}. The intersection of these two domains of physics holds promising potential for unveiling deeper insights into the fundamental nature of the universe.

Quantum field theory has achieved remarkable success in deciphering the complexities of interactions among subatomic particles and in shedding light on the origins of the three fundamental forces in nature: the weak, strong, and electromagnetic interactions \citep{key-5}. Nonetheless, the pursuit of a unified theory capable of reconciling general relativity and quantum mechanics, often referred to as a theory of quantum gravity, has long been beset by persistent challenges and unresolved technical issues, at least until very recently \citep{key-6,key-7}. These hurdles have spurred intense scientific endeavors aimed at bridging the lingering gaps in our understanding, as scientists work diligently to unveil the foundational framework that harmonizes these two fundamental cornerstones of modern physics.

Som and Raychaudhuri \citep{key-07} introduced a set of stationary cylindrically symmetric solutions to the Einstein–Maxwell equations, which describe a charged dust distribution undergoing rigid rotation. In this solution, the Lorentz force becomes zero everywhere, and the ratio of charge density to mass density can take on any arbitrary value. Barrow and Dabrowski \citep{key-08} demonstrated that in low-energy effective string theories, homogeneous Gödel-type space-times may not contain closed time-like curves. They presented exact solutions for the Gödel-type metric within string theory \citep{key-09}, considering the full $ O\left(\alpha\right)$ action, which includes both dilaton and axion fields \citep{key-08,key-09,key-010}. It's worth noting that the Som-Raychaudhuri space-time belongs to the category of flat Gödel-type solutions. Various aspects of Gödel-type solutions are elaborated upon in \citep{key-011,key-012,key-013,key-014}. Gürses et al. illustrated through several examples that Gödel-type metrics can be applied to derive exact solutions in various supergravity theories, resulting in spacetimes that may encompass closed time-like and closed null curves or none of these features \citep{key-015}. Furthermore, Clifton et al. \citep{key-016} identified the general conditions for the existence of Gödel, Einstein static, and de-Sitter universes in gravity theories that are derived from a Lagrangian defined as an arbitrary function of the scalar curvature and Riemann curvature invariants. They provided explicit expressions for these solutions in terms of the Lagrangian parameters and determined the conditions under which time travel is permissible in Gödel universes.

Numerous research studies have probed the examination of physical attributes associated with various backgrounds in the context of Som Raychaudhuri spacetime. These investigations encompass a wide array of phenomena, including: (i) The characteristics of the rotating Som–Raychaudhuri homogeneous spacetime. (ii) The relativistic quantum dynamics of spin-zero particles in Som–Raychaudhuri spacetime under the gravitational influence of a specific topology \citep{key-017}. (iii) The behavior of the Klein–Gordon oscillator in Som–Raychaudhuri spacetime within a cosmic context \citep{key-018}. (iv) The confinement of scalar particles in a linear fashion within Som–Raychaudhuri spacetime \citep{key-019}, and (v) The confinement of scalar particles in a topologically straightforward flat Gödel-type spacetime \citep{key-020}.

Additionally, another reference, Ref. \citep{key-021}, explores the quantum dynamics of scalar and spin-half particles in Gödel-type spacetimes characterized by positive, negative, and zero curvatures. In Refs. \citep{key-022,key-023}, the quantum dynamics of scalar particles in a particular category of Gödel-type solutions are examined, revealing significant similarities in energy levels to the Landau problem within flat, spherical, and hyperbolic spaces. In Ref. \citep{key-024}, the investigation focuses on the quantum dynamics of scalar particles within Som–Raychaudhuri spacetime, offering a comparison with Landau levels in flat space. Ref. \citep{key-025} delves into the quantum dynamics of spin-half particles (Dirac fermions) within the backdrop of Som–Raychaudhuri spacetime, featuring the presence of torsion and a cosmic string traversing through it. The researchers observe a break in the degeneracy of relativistic energy levels, and they find that the corresponding eigenfunctions are influenced by the presence of the topological defect in the background spacetime with torsion. Ref. \citep{key-026} investigates the quantum dynamics of spin-half particles (Weyl fermions) in Som–Raychaudhuri spacetime featuring a topological defect, while in Ref. \citep{key-027}, the study explores scalar quantum particles within a specific class of Gödel-type solutions with a cosmic string intersecting the spacetime. Ref. \citep{key-028} is dedicated to the investigation of the Fermi field and Dirac oscillator within the context of Som–Raychaudhuri spacetime.

In addition, Ref. \citep{key-029} examines the Klein–Gordon equation when subjected to vector and scalar potentials of Coulomb-type under the influence of non-inertial effects in the cosmic string spacetime. Finally, Ref. \citep{key-030} investigates two distinct categories of solutions for the Klein–Gordon equation in the presence of Coulomb-like scalar potentials while considering non-inertial effects in the cosmic string spacetime.

This paper provides a comprehensive investigation into the dynamic behavior of boson particles in the presence of a cosmic string in Som Raychaudhuri spacetime. To do this we remark that: (i) we focuses on studying the effects of a cosmic string on the free Klein-Gordon equation in Som Raychaudhuri spacetime, analyzing how the equation is modified and its implications on the Klein-Gordon field dynamics. (ii) Then, we extends its inquiry to the Klein-Gordon oscillator in this spacetime context, providing an exact solution and exploring the resulting energy spectrum. This comprehensive investigation enhances our understanding of the Klein-Gordon equation's behavior in the presence of cosmic strings, offering insights into the quantized energy levels and their implications for the universe's dynamics, bridging theory and observation in relativistic wave equations' interactions with cosmic strings.

The outline of this paper is as follows: Section II is devoted as an overview of the Klein-Gordon equation in curved spacetime. The solution of free and Klein-Gordon oscillator are the main objects of both  Sections III and IV. Finally, we give conclusions in Section V.

\section{An Overview of the Klein-Gordon equation}

In this section, our objective is to examine the Klein-Gordon oscillator
(KGO) within the framework of a cosmic string background geometry
. It is widely acknowledged that the relativistic wave equations for
a scalar particle in a Riemannian spacetime, defined by the metric
tensor $g_{\alpha\beta}$ , can be derived by reformulating the Klein-Gordon
equation in a manner described in textbooks such as \citep{key-8,key-9,key-10,key-0101,key-0102,key-0103}.
\begin{equation}
\left(\frac{1}{\sqrt{-g}}\partial_{\alpha}\left(\sqrt{-g}g^{\alpha\beta}\partial_{\beta}\right)-m^{2}-\xi R\right)\boldsymbol{\phi}(\boldsymbol{r},t)=0,\label{eq:1}
\end{equation}
Here, we have the Laplace-Beltrami operator $\square=\frac{1}{\sqrt{-g}}\partial_{\alpha}\left(\sqrt{-g}g^{\alpha\beta}\partial_{\beta}\right)$denoted
as $\nabla^{2}$ , $\xi$ represents a real dimensionless coupling
constant, and $R$ is the Ricci scalar curvature defined as $R=g^{\alpha\beta}R_{\alpha\beta}$
, where $R_{\mu\nu}$ denotes the Ricci curvature tensor. $g^{\alpha\beta}$
represents the inverse metric tensor, and $g=\det(g_{\mu\nu})$.

Our objective is to investigate the quantum dynamics of spin-0 particles
in the spacetime induced by the Som-Raychaudhuri spacetime with cosmic
string oscillators.

\section{Free Klein-Gordon equation in Som\textendash Raychaudhuri spacetime
with a cosmic string }

Prior to delving into the Hamiltonian representation of the KGO, let
us commence by deriving the Klein-Gordon (KG) wave equation for a
free relativistic scalar particle in the context of a cosmic string
spacetime. This particular spacetime is presumed to be static and
possess cylindrical symmetry.

The line element that characterizes the cosmic string metric in a
(1+3)-dimensional Som-Raychaudhuri spacetime (with $\hbar=c=1$) is
given by \citep{key-11,key-12,key-13,key-14}:
\begin{align}
ds^{2} & =g_{\mu\nu}dx^{\mu}dx^{\nu}\nonumber \\
 & =-\left(dt+\alpha\Omega r^{2}d\varphi\right)^{2}+dr^{2}+\alpha^{2}r^{2}d\varphi^{2}+dz^{2},\label{eq:2}
\end{align}
while the parameter $\alpha$ characterizes the cosmic string with
$\alpha\in\left[0,1\right]$ , The parameter $\Omega$ characterizes
the vorticity of the spacetime where
\begin{equation}
ds^{2}=-dt^{2}-2\Omega\alpha r^{2}dtd\varphi+dr^{2}+\left[\alpha^{2}r^{2}-\left(\alpha\varOmega r^{2}\right)^{2}\right]d\varphi^{2}+dz^{2},\label{eq:3}
\end{equation}
Given the metric and inverse metric tensor components, represented
as follows:
\begin{equation}
g_{\mu\nu}=\left(\begin{array}{cccc}
-1 & 0 & -\alpha\Omega r^{2} & 0\\
0 & 1 & 0 & 0\\
-\alpha\Omega r^{2} & 0 & \left[\alpha^{2}r^{2}-\left(\alpha\varOmega r^{2}\right)^{2}\right] & 0\\
0 & 0 & 0 & 1
\end{array}\right),\quad g^{\mu\nu}=\left(\begin{array}{cccc}
\left(-1+\alpha\left(\Omega r\right)^{2}\right) & 0 & -\Omega & 0\\
0 & 1 & 0 & 0\\
-\Omega & 0 & \frac{1}{\left(\alpha r\right)^{2}} & 0\\
0 & 0 & 0 & 1
\end{array}\right)\label{eq:4}
\end{equation}
After performing some straightforward algebraic manipulations, we
obtain the following second-order differential equation for the radial
function $\psi(r)$:
\begin{equation}
\left[\left(-1+\alpha\left(\Omega r\right)^{2}\right)\frac{\partial^{2}}{\partial t^{2}}+\frac{1}{\sqrt{-g}}\frac{\partial}{\partial r}\left(\sqrt{-g}\frac{\partial}{\partial r}\right)-\frac{1}{\alpha^{2}r^{2}}\frac{\partial^{2}}{\partial\varphi^{2}}+\frac{\partial^{2}}{\partial z^{2}}-2\varOmega\left(\frac{\partial}{\partial t}\frac{\partial}{\partial\varphi}\right)-m^{2}\right]\boldsymbol{\phi}\left(\boldsymbol{r}\right)=0\label{eq:5}
\end{equation}
The wave function of this system is defined by
\begin{equation}
\boldsymbol{\phi}\left(\boldsymbol{r}\right)=e^{-iEt+i\ell\varphi+iKz}\psi\left(r\right)\label{eq:6}
\end{equation}
Where $\ell=0,\pm1,\pm2,\pm3....$. By substituting the last equation
into the second-order differential equation, we obtain:
\begin{equation}
\left[\frac{d^{2}}{dr^{2}}+\frac{1}{r}\frac{d}{dr}-\frac{\left(\ell\right)^{2}}{\alpha^{2}r^{2}}-\alpha\left(\Omega E\right)^{2}r^{2}+E^{2}-m^{2}-K^{2}-2\ell\varOmega E\right]\psi\left(r\right)=0\label{eq:7}
\end{equation}
setting $\gamma=E^{2}-m^{2}-K^{2}-2\ell\varOmega E,$ yields
\begin{equation}
\left[\frac{d^{2}}{dr^{2}}+\frac{1}{r}\frac{d}{dr}-\frac{\ell^{2}}{\alpha^{2}r^{2}}-\alpha\left(\Omega E\right)^{2}r^{2}+\gamma\right]\psi\left(r\right)=0,\label{eq:8}
\end{equation}
{\bf{Observing Eq. (\ref{eq:8}) we can note that, although the particle is considered "free", in fact, in the radial direction it is subject to a harmonic-type central potential $(-\alpha\left(\Omega E\right)^{2}r^{2})$. It is worth mentioning that this potential has a gravitational nature, that is, it was induced by the background characterized by the metric given in Eq. (\ref{eq:3}).}} The exact solution of this The exact solutions to this differential equation are given by Whittaker functions \cite{key-14a}, where:
\begin{align}
\psi\left(r\right) & =\mathcal{C}_{1}r^{-1}\mathrm{WhittakerM}\left(\frac{\gamma}{4\left(\Omega E\alpha^{1/2}\right)},\frac{\ell}{2\alpha},\alpha^{1/2}\Omega E\,r^{2}\right)\nonumber \\
 & +\mathcal{C}_{2}r^{-1}\mathrm{WhittakerW}\left(\frac{\gamma}{4\left(\Omega E\alpha^{1/2}\right)},\frac{\ell}{2\alpha},\alpha^{1/2}\Omega E\,r^{2}\right)\label{eq:9}
\end{align}
by using this condition of the wave function $\psi\left(0\right)=0,$
and $\underset{r\rightarrow+\infty}{Lim}\psi\left(r\right)=0$, suggest
the exact wave function of the system,where 
\begin{equation}
\psi\left(r\right)=\mathcal{N}_{1}r^{-1}\mathrm{WhittakerM}\left(\frac{\gamma}{4\left(\Omega E\alpha^{1/2}\right)},\frac{\ell}{2\alpha},\alpha^{1/2}\Omega E\,r^{2}\right)\label{eq:10}
\end{equation}
where $\mathcal{N}_{1}$ represents the normalization condition, The
relationship between Whittaker functions $WhittakerM\left(a,b,z\right)$
and hypergeometric polynomials $_{1}F_{1}\left(a,b,z\right)$ is given by \cite{key-14b}:
\begin{equation}
WhittakerM\left(a,b,z\right)=z^{b+\frac{1}{2}}e^{-z/2}{}_{1}F_{1}\left(\frac{1}{2}-a+b,2b+1,z\right)\label{eq:11}
\end{equation}
The final relation of the wave function with the hypergeometric polynomial
is given as:
\begin{equation}
\psi\left(r\right)=\mathcal{N}_{1}r^{-1}\mathrm{e}^{\frac{\Omega E\alpha^{\frac{1}{2}}\,r^{2}}{2}}\left(\alpha^{\frac{1}{2}}\Omega E\,r^{2}\right){}^{\frac{\alpha+\ell}{2\alpha}}\mathrm{_{1}F_{1}}\left(-\frac{-2E\Omega\,\alpha^{\frac{3}{2}}-2E\Omega\ell\alpha^{\frac{1}{2}}+\alpha\left(-2\ell E\Omega+E^{2}-K^{2}-m^{2}\right)}{4\alpha^{\frac{3}{2}}E\Omega},1+\frac{\ell}{\alpha},\alpha^{\frac{1}{2}}\Omega E\,r^{2}\right)\label{eq:12}
\end{equation}
We should note that in equation \eqref{eq:12}, the solution must
be a polynomial function of degree $n$. However, when $n\to\infty$,
a divergence issue arises. We can only have a finite polynomial if
the factor of the first term in equation \eqref{eq:12} is a negative
integer, indicating that:
\begin{equation}
\frac{2E\Omega\,\alpha^{\frac{3}{2}}+2E\Omega\ell\alpha^{\frac{1}{2}}+\alpha\left(2\ell E\Omega-E^{2}+K^{2}+m^{2}\right)}{4\alpha^{\frac{3}{2}}E\Omega}=-n\label{eq:13}
\end{equation}
By utilizing this result and substituting the parameters from equation
\eqref{eq:13}, we can determine the quantized energy spectrum of
the free Klein-Gordon equation within the Som-Raychaudhuri spacetime
of the cosmic string. Therefore, we obtain:
\begin{align}
E_{n}^{\pm} & =\pm\Omega\left[\left(2n+1\right)\,\alpha^{\frac{1}{2}}+\ell\left(1+\alpha^{-\frac{1}{2}}\right)\right]\\
 & \pm\sqrt{\frac{\left(4\Omega^{2}\ell\left(n+\frac{1}{2}\right)\alpha^{\frac{3}{2}}+2\alpha^{\frac{1}{2}}\,\Omega^{2}\ell^{2}+\left(4\left(n+\frac{1}{2}\right)^{2}\alpha^{2}+\left(\ell^{2}+\left(4n+2\right)\ell\right)\alpha+\ell^{2}\right)\Omega^{2}+\alpha\left[m^{2}+K^{2}\right]\right)}{\alpha}} \label{eq:14}
\end{align}
Equation (\ref{eq:14}) displays the energy spectrum of the free Klein-Gordon equation in the context of the Som-Raychaudhuri spacetime. The form of this equation is dependent on different geometrical parameters of the spacetime. To understand the effect of these parameters on the energy levels, figures. \ref{Fig free 1} and  \ref{Fig free 2} have been constructed for this purpose. Based on the two figures, we can  observed the following :

\begin{center}
\begin{figure}
\subfloat[here $\ell=1$]{\centering{}\includegraphics[scale=0.65]{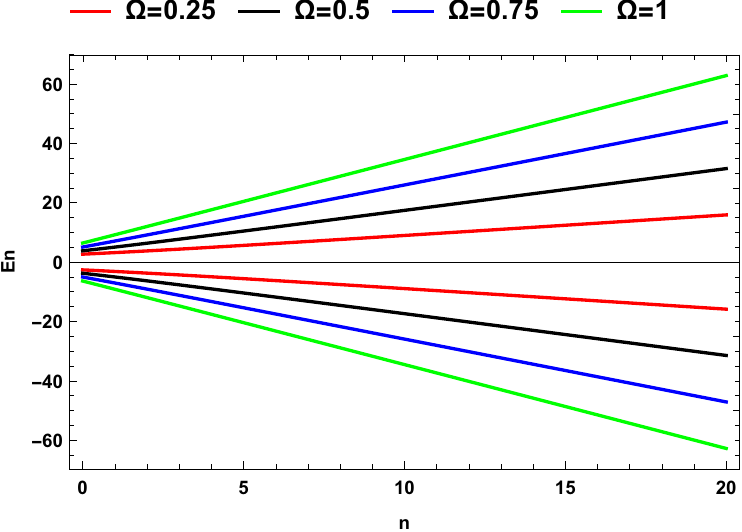}}\subfloat[here $\ell=0$]{\centering{}\includegraphics[scale=0.65]{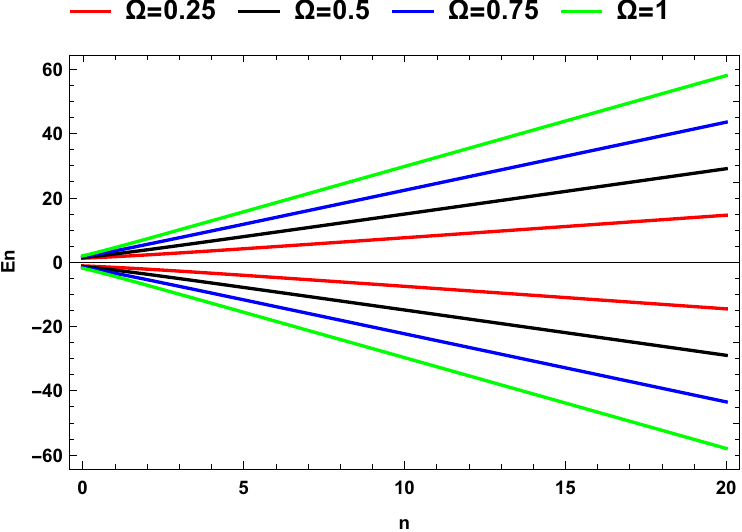}}\par
\caption{Spectrum of energy versus quantum number $n$ for different values of $\varOmega$ and $\ell=0,1$. here $m=\omega=1,\alpha=0.5,K=0$ }
\label{Fig free 1}
\end{figure}
\begin{figure}
\subfloat[here $\ell=1$]{\centering{}\includegraphics[scale=0.65]{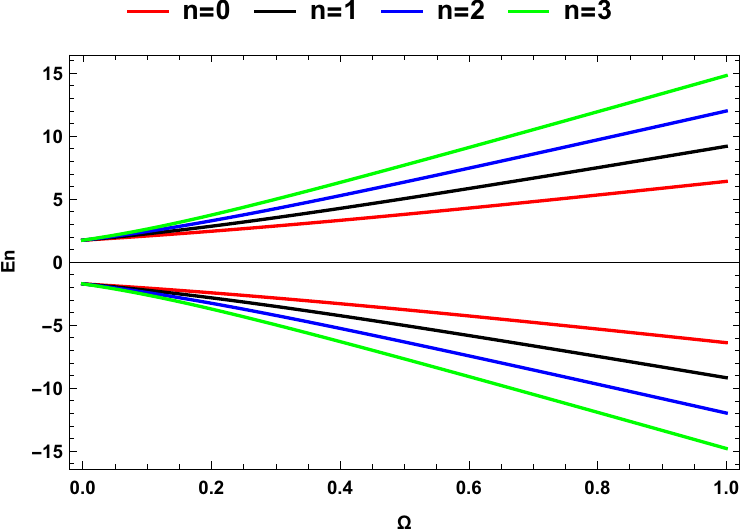}}\subfloat[here $\ell=0$]{\centering{}\includegraphics[scale=0.65]{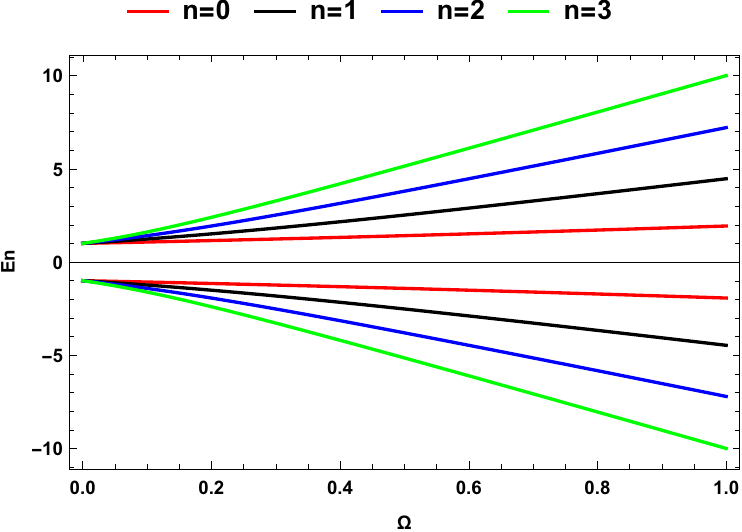}}\par
\caption{Spectrum of energy versus $\varOmega$ for four level $n=0,1,2,3$ and $\ell=0,1$. here $m=\omega=1,\alpha=0.5,K=0$ }
\label{Fig free 2}
\end{figure}\par
\end{center}
\begin{center}
\begin{figure}[H]
\begin{center}
\subfloat[here $\ell=1$]{\centering{}\includegraphics[scale=0.65]{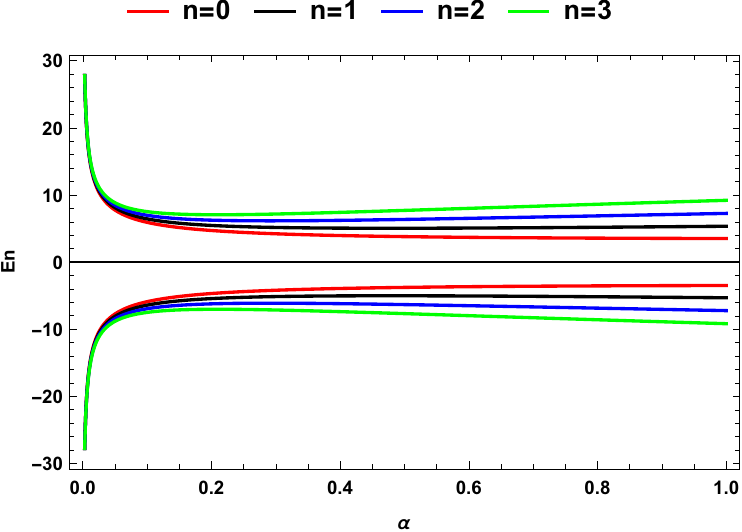}}\subfloat[here $\ell=0$]{\centering{}\includegraphics[scale=0.65]{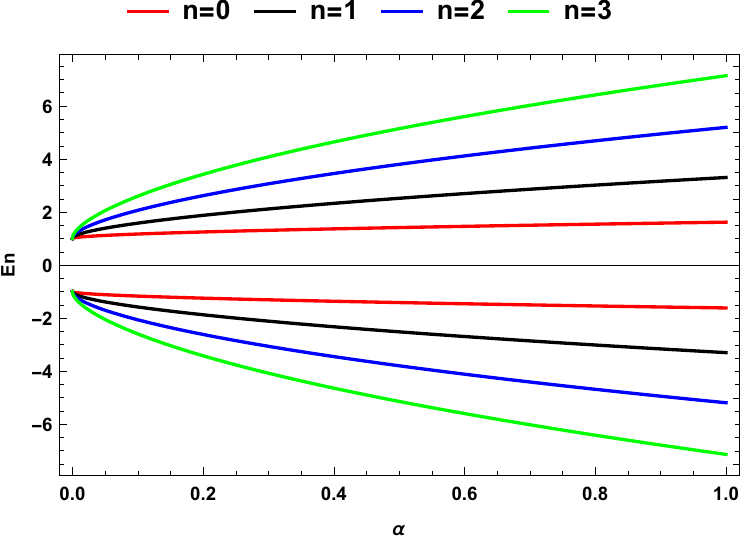}}\par
\caption{Spectrum of energy versus $\alpha$ for four level $n=0,1,2,3$ and $\ell=0,1$. here $m=\omega=1,\alpha=0.5,K=0$ }
\label{Fig free 3}
\end{center}
\end{figure}
\par\end{center}
In Figure \ref{Fig free 1}, we  illustrate the energy levels of the free Klein-Gordon equation when operating within the framework of Some-Raychaudhuri spacetime in function of the quantum number $n$ . In this investigation, we have standardized certain key parameters: both mass ($m$) and frequency ($w$) are set to a value of 1, while the parameters $\alpha=0.5$ and $K=0$. We have systematically explored various scenarios, altering the value of the parameter $\Omega$. Two specific cases have been carefully examined: in figure(1.a),  the angular momentum $\ell=1$ , while in figure(1.b), $\ell=0$.  Our primary goal has been to gain a deep understanding of how the $\Omega$ term influences the structure of the spacetime geometry and to discern its effectiveness in shaping the dynamic behavior of energy within the system. These findings have significant implications for advancing our comprehension of how the motion of particles behave in the context of curved spacetime.

Figure \ref{Fig free 2} present the energy spectrum of the free Klein-Gordon equation within the framework of Some-Raychaudhuri spacetime for different values of $\varOmega$ using the same adjustable parameters. This study encompasses a thorough exploration of various quantum number values $n=0,1,2,3$) within different visions. Specifically, we investigate two cases: case  where the orbital angular momentum quantum number ($\ell$) is set to 1, and the other with $\ell=0$. Our inquiry extends across the spectrum of geometrical parameter values within the interval [0, 1], focusing on the impact of the $\Omega$ term in relation to the spacetime geometry. Our findings illuminate that the influence of the $\Omega$ term on the geometric spacetime structure is consistently effective across all quantum number levels, thus underscoring its significance in shaping the energy spectrum of the Klein-Gordon equation in Some-Raychaudhuri spacetime.

Finally, Figure \ref{Fig free 3} depict the Spectrum of energy versus $\alpha$ for four level $n=0,1,2,3$ and $\ell=0,1$ within the context of Some-Raychaudhuri spacetime.  two distinct scenarios can be study: case 'a,' where the angular momentum ($\ell=1$) and the other where $\ell=0$. While we explore different quantum number levels ($n=0,1,2,3$), a conspicuous influence of the $\alpha$ term emerges consistently across all these levels. In the instance of $\ell=1$ and $\alpha=0$, a striking observation surfaces, where the energy levels approach both positive and negative infinity, indicating the presence of a pole in the energy spectrum. This phenomenon signifies the crucial role that the topological defect plays in modifying energy levels when $\ell=1$. In contrast, for $\ell=0$, we observe a symmetrical energy spectrum, with the impact of the topological defect maintaining a high degree of effectiveness, continuously influencing energy levels. These remarks shed light on the profound impact of topological defects on the dynamic behavior of energy within the Klein-Gordon equation, highlighting their intricate interaction with spacetime geometry and quantum properties.

In conclusion, the analysis of the energy spectrum of the free Klein-Gordon equation within Some-Raychaudhuri spacetime leads to a significant revelation: the quantization of energy levels, expressed as $E^{\pm}=f(n)$. This quantization emphasizes the pivotal role played by the geometrical parameter defined in the initial term of the inverse metric component $g^{00}$. The significance of this observation lies in its direct correlation to the underlying spacetime geometry. The inverse metric component $g^{00}$ captures the gravitational field's influence on the energy of Klein-Gordon particles, rendering it an indispensable factor in our comprehension of how gravitation affects the quantum behavior of particles within this specific spacetime.
As a consequance, this finding signifies a profound connection between spacetime geometry and the quantized energy levels of particles, unveiling an intriguing link between general relativity and quantum mechanics within the framework of the Klein-Gordon equation in Some-Raychaudhuri spacetime. In addition, These observations illuminate the intricate relationship between particle dynamics and the underlying spacetime geometry. The parameter $\Omega$ plays a pivotal role in shaping this interaction, emphasizing the profound interplay between the fundamental constituents of the universe and the fabric of this spacetime. in the case of $\left(\Omega=0\right)$, the spectrum of the free Klein-Gordon equation within the Som-Raychaudhuri spacetime of the
cosmic string is given by $E^{\pm}=\pm\sqrt{m^{2}+K^{2}}$.

In the next section of this investigation, we will introduce the the interaction of the Dirac oscillator.

\section{Klein-Gordon oscillator in Som\textendash Raychaudhuri spacetime
with a cosmic string}
In this section, our interest is to describe the quantum dynamics of a relativistic oscillator model in spacetime described by the metric given in Eq. (\ref{eq:3}). This oscillator model is known in the literature as the Klein-Gordon oscillator (KGO) \cite{kgo}, which has become an object of study in several branches of physics due to its unique characteristics that other models of relativistic oscillators do not have. Among these characteristics is analyticity, that is, it is possible to describe it completely analytically, without needing numerical or approximate methods to obtain its particular solution or chaos. Another feature that makes it very interesting is the recovery of the Schrondiger oscillator in the non-relativistic regime \cite{kgo1}.

KGO has been investigated in several relativistic quantum mechanics scenarios, for example, subjected to the central potentials in Minkowski spacetime \cite{kgo2,kgo3,kgo4}, in cosmic string spacetime \cite{kgo5}, in global monopole spacetime \cite{kgo6}, in Kaluza-Klein theory \cite{kgo7,kgo8,kgo9}, and in possible scenarios of Lorentz symmetry violation \cite{kgo10,kgo11}.

We begin by considering a scalar quantum particle residing in the
gravitational field of the spacetime described by the metric \eqref{eq:2}.
To incorporate this, we introduce a modification of the momentum operator,
replacing $p_{i}$ with $p_{i}\longrightarrow p_{i}+im\omega x_{i}$
, where $p_{i}=i\nabla_{i}$ in Eq \eqref{eq:8}.

By following the same steps as before, we derive the following radial
equation:
\begin{equation}
\left[\frac{d^{2}}{dr^{2}}+\frac{1}{r}\frac{d}{dr}-\frac{\ell^{2}}{\alpha^{2}r^{2}}-\left[m^{2}\omega^{2}+\alpha\left(\Omega E\right)^{2}\right]r^{2}+\delta\right]\psi\left(r\right)=0,\label{eq:15}
\end{equation}

By observing Eq. (\ref{eq:15}) we can note that that the oscillator term remains, however, this term is effected by the term arising from the non-minimal coupling in the Klein-Gordon equation, which guarantees the presence of the KGO.

Here, we have defined:
\begin{equation}
\delta=E^{2}-m^{2}-K^{2}-2\ell\varOmega E+2m\omega.\label{eq:16}
\end{equation}
Following mathematical computations, it becomes possible to present the precise expression of the wave function for the analyzed system.
\begin{align}
\psi\left(r\right) & =\mathcal{C}_{1}r^{-1}\mathrm{WhittakerM}\left(\frac{\delta}{4\sqrt{\alpha E^{2}\Omega^{2}+m^{2}\omega^{2}}},\frac{\ell}{2\alpha},\sqrt{\alpha E^{2}\Omega^{2}+m^{2}\omega^{2}}\,r^{2}\right)\nonumber \\
 & +\mathcal{C}_{2}r^{-1}\mathrm{WhittakerW}\left(\frac{\delta}{4\sqrt{\alpha E^{2}\Omega^{2}+m^{2}\omega^{2}}},\frac{\ell}{2\alpha},\sqrt{\alpha E^{2}\Omega^{2}+m^{2}\omega^{2}}\,r^{2}\right)\label{eq:17}
\end{align}
By applying convergence criteria to the previously utilized wave functions, it is possible to evaluate the following equation as a potential wave function.
\begin{equation}
\psi\left(r\right)=\mathcal{N}_{2}\mathrm{WhittakerM}\left(\frac{\delta}{4\sqrt{\alpha E^{2}\Omega^{2}+m^{2}\omega^{2}}},\frac{\ell}{2\alpha},\sqrt{\alpha E^{2}\Omega^{2}+m^{2}\omega^{2}}\,r^{2}\right)\label{eq:18}
\end{equation}
The primary goal of mathematical approximations is to consistently discover actual solutions that elucidate concealed physical phenomena. In the realm of scientific inquiry, the ability to accurately model and understand the behavior of complex systems is of paramount importance. By employing mathematical conversions, such as transitioning from Whittaker functions to hypergeometric functions, researchers can navigate the intricate landscape of quantum mechanics, particle physics, and other branches of scientific study. These conversions serve as powerful tools that allow for a deeper exploration of the underlying principles governing the universe. They provide a means to bridge the gap between abstract mathematical formulations and tangible observations, facilitating the interpretation and prediction of experimental results. As these conversions unfold, revealing the intricate relationship between various mathematical constructs, researchers gain valuable insights into the fundamental nature of the systems under investigation. Through this process, new connections and patterns emerge, shedding light on previously uncharted aspects of the physical world and paving the way for further scientific advancements. Thus, the application of mathematical conversions, particularly the transition from Whittaker to hypergeometric functions, serves as a gateway to unraveling the mysteries that lie hidden within the fabric of our reality.
\begin{align}
\psi\left(r\right) & =\mathcal{N}_{2}r^{-1}\mathrm{e}^{-\frac{\sqrt{\alpha E^{2}\Omega^{2}+m^{2}\omega^{2}}\,r^{2}}{2}}\left(\sqrt{\alpha E^{2}\Omega^{2}+m^{2}\omega^{2}}\,r^{2}\right){}^{\frac{\alpha+\ell}{2\alpha}}\nonumber \\
 & \mathrm{_{1}F_{1}}\left(-\frac{(-2\ell-2\alpha)\sqrt{\alpha E^{2}\Omega^{2}+m^{2}\omega^{2}}+(-2\ell\Omega E+E^{2}-K^{2}-m^{2}+2m\omega)\alpha}{4\sqrt{\alpha E^{2}\Omega^{2}+m^{2}\omega^{2}}\,\alpha},1+\frac{\ell}{\alpha},\sqrt{\alpha E^{2}\Omega^{2}+m^{2}\omega^{2}}\,r^{2}\right)\label{eq:19}
\end{align}
In the quest for a comprehensive understanding of the studied system, the culmination of mathematical approximations leads us to the solution of the final equation for the wave functions. This solution is expressed in terms of hypergeometric functions, which play a vital role in capturing the intricate behavior of quantum systems. The hypergeometric function, with its powerful mathematical properties, enables us to unravel the complexities of the system and explore its underlying dynamics in greater detail. By providing a precise and elegant mathematical representation, this solution unveils the intricate relationships between the various elements of the system, shedding light on the hidden physical phenomena at play. Through this remarkable feat of mathematical analysis, we are able to grasp a deeper understanding of the system's behavior and uncover profound insights into the nature of the spectrum of energy.
Again, The asymptotic behavior of the confluent hypergeometric function
implies that
\begin{equation}
\frac{E^{2}-m^{2}-K^{2}-2\ell\varOmega E+2m\omega}{4\sqrt{\left(\alpha E^{2}\Omega^{2}+m^{2}\omega^{2}\right)}}=-n,\label{eq:20}
\end{equation}
Indeed, it is acknowledged that the equation at hand lacks an exact and explicit solution due to its inherent complexity and difficulty. Such intricate equations often pose significant challenges when seeking closed-form solutions. However, to facilitate a deeper understanding and simplify the analysis, we can focus on obtaining the exact form of the system's energy. Remarkably, the energy expression is derived and presented in the form of a fourth-degree equation. While the complexity remains, this formulation provides a valuable stepping stone for further investigation and analysis. By examining the energy equation, we can discern critical characteristics and properties of the system's energy landscape, allowing us to gain insights into the system's behavior, stability, and possible energy states. While the equation may present mathematical and computational challenges, its exact form grants us an invaluable starting point for further exploration and deeper comprehension of the underlying physical phenomena.

After streamlining the mathematical calculations, we successfully unveil the energy spectrum of the system, representing a significant milestone in our analysis. Through meticulous simplifications and computations, we arrive at an explicit form that characterizes the system energy states. The energy spectrum emerges in a distinct and compelling fashion, offering valuable insights into the system's behavior and the quantized nature of its energy levels where  
\begin{equation}
\alpha^{2}E^{4}-4\ell\Omega\alpha^{2}E^{3}-aE^{2}+bE-c=0
\label{eq:energy}
\end{equation}
with
\begin{align*}
a & =16\Omega^{2}\alpha^{3}n^{2}+16\Omega^{2}\alpha^{3}n-4\Omega^{2}\alpha^{2}\ell^{2}+16\Omega^{2}\alpha^{2}\ell n+4\Omega^{2}\alpha^{3}+8\Omega^{2}\alpha^{2}\ell+4\Omega^{2}\alpha\,\ell^{2}+2K^{2}\alpha^{2}+2\alpha^{2}m^{2}-4\alpha^{2}m\omega\\
b & =4\ell\Omega\alpha^{2}\left[m^{2}+K^{2}\right]-8\Omega\,\alpha^{2}\ell m\omega\\
c & =16\alpha^{2}m^{2}n^{2}\omega^{2}+16\alpha^{2}m^{2}n\omega^{2}+16\alpha\ell m^{2}n\omega^{2}-K^{4}\alpha^{2}-2K^{2}\alpha^{2}m^{2}+4K^{2}\alpha^{2}m\omega-\alpha^{2}m^{4}+4\alpha^{2}m^{3}\omega+8\alpha\ell m^{2}\omega^{2}+4\ell^{2}m^{2}\omega^{2}
\end{align*}
The energy spectrum $E^{\pm}\left(n\right)$ (Eq. (\ref{eq:energy})is an algebraic equation of the degree 4 having both real and complex solutions. The complex solutions are not physical, but there are two other real solutions. The energy spectrum will be discussed for different values of $\varOmega$ and $\ell$.
To better understand the behavior of our energy spectrum, we can consider two situations. The first one is the case when  $\varOmega=0$, and the second is when $\varOmega \neq0$.

For the first case when  $\varOmega=0$, it represents a special case that allows us to derive the energy spectrum associated with the extensively researched findings by Boumali and Messai \citep{kgo5}. In their comprehensive investigation, they delved into how the topological defect constant, denoted as $\alpha$, influences the energy spectrum of cosmic strings. The topological defect constant $\alpha$ is a pivotal factor in determining the resulting energy spectrum, with its definition stemming from the physical attributes and geometric configuration of cosmic strings. The outcomes of their research underscore the substantial impact of $\alpha$ values on the energy distribution. Specifically, when $\omega$ is set to zero, the resulting findings align with the extensively explored spectrum outlined in reference \citep{kgo5}.

In addition, their study contributes significantly to advancing our comprehension of the intricate interplay between topological defect constants and the energy spectrum of cosmic strings. By delving into the characteristics and properties of these topological defects, researchers can uncover the fundamental mechanisms governing the formation, evolution, and behavior of cosmic strings. The energy spectrum derived in the special scenario where $\Omega=0$ holds particular importance as it provides insights into the specific dynamics and quantum nature of cosmic strings when other contributing factors are absent. where:
\begin{equation}
E^{\pm}\left(n\right)=\pm\sqrt{\left(4nm\omega+\frac{2m\omega\left|\ell\right|}{\alpha}+m^{2}+K^{2}\right)}\label{eq:22}
\end{equation}

Now, let's examine the scenario where $\varOmega$ is not equal to 0, and in this context, we can distinguish between two cases.
\begin{itemize}
  \item The first Scenario (When $\ell = 0$): In this case, the numerical values of the spectrum of energy are given into Table. 1. This table provides a comprehensive summary of the energy levels within the system under investigation, encompassing quantum numbers within the range of $[0, 21]$, with a fixed value of the parameter $\alpha$ set to $0.5$. The obtained results  enable a comprehensive exploration of the system's quantum characteristics. The associated values have been plotted for various $\varOmega$ values in Figure. \ref{fig.4}. As shown in this figure, the energy spectrum displays symmetry and increases with growing $\varOmega$.
 \item The second Scenario (With $\ell = 1$): In this  scenario,  specific numerical values of the energy spectrum, as discussed in the previous section, are reported in Table 2. These values have been visualized in Figure. \ref{fig.5} for various $\varOmega$ values, while maintaining a fixed value of the parameter $\alpha$ at $0.5$. Contrarily the previous case, the energy spectrum in this scenario exhibits asymmetry.
\end{itemize}
In summary, this research paper's thorough investigation significantly enhances our comprehension of how the Klein-Gordon equation behaves within the context of Som Raychaudhuri spacetime when influenced by a cosmic string. The findings obtained through this study contribute to a deeper understanding of relativistic wave equations and their interactions with cosmic structures. Moreover, the insights garnered from this research have broad-reaching implications for advancing our understanding of the universe's dynamics, encompassing both theoretical progress and potential links to observable phenomena.

Moving forward, we plan to replicate this study with varying components and explore different mathematical approximations. Additionally, we intend to investigate the thermal characteristics of the materials involved and employ the concept of generalization to develop more precise approximation models, drawing upon the Feshbach-Villars approach.

\section{Conclusion }

"This paper offers a comprehensive examination of the behavior of the Klein-Gordon equation within the context of Som Raychaudhuri spacetime when influenced by a cosmic string. We establish the fundamental significance of the Klein-Gordon equation in theoretical physics, emphasizing its role as a foundational framework for describing relativistic wave equations. This foundational understanding forms the basis for our exploration of the equation's behavior when cosmic strings are introduced.

Subsequently, we delve into the modified behavior of the free Klein-Gordon equation when a cosmic string is present in Som Raychaudhuri spacetime. This analysis takes into account the effects of the cosmic string, revealing insights into the alterations in the equation's dynamics and their implications for the behavior of the Klein-Gordon field.

In the final phase of our study, we extend our investigation to the Klein-Gordon oscillator within Som Raychaudhuri spacetime with a cosmic string. Through the examination of an exact solution, our research delves into the energy spectrum associated with this system. These inquiries provide valuable insights into the quantization of energy levels, shedding light on the dynamics of the Klein-Gordon oscillator."

\newpage
\begin{table}[h]
\caption{Some  values of energies  for  $\ell=0,m=\omega=1,\alpha=0.5,K=0$}
\centering{}{}%
\begin{tabular}{ccccccccc}
\hline 
\hline
\textbf{Quantum number, $n$} & \multicolumn{2}{|c|}{{$\Omega=0.25$}} & \multicolumn{2}{c|}{{$\Omega=0.5$}} & \multicolumn{2}{c|}{{$\Omega=0.75$}} & \multicolumn{2}{c|}{{$\Omega=1$}}
\tabularnewline
\hline
\hline 
\multicolumn{1}{c}{
\begin{turn}{} \end{turn}} & {$E^{+}$} & {$E^{-}$} & {$E^{+}$} & {$E^{-}$} & {$E^{+}$} & {$E^{-}$} & {$E^{+}$} & {$E^{-}$}
\tabularnewline
{0} & {01.01587} & {-01.01587} & {1.06651} & {-01.06651} & {01.16144} & {-01.16144} & {01.31607} & {-01.31607}
\tabularnewline

{1} & {01.98577} & {-01.98577} & {2.40511} & {-02.40511} & {03.16559} & {-03.16559} & {04.13647} & {-04.13647}
\tabularnewline
 
{2} & {02.96094} & {-02.96094} & {3.86966} & {-03.86966} & {05.33645} & {-05.33645} & {07.02548} & {-07.02548}
\tabularnewline
 
{3} & {03.78960} & {-03.78960} & {5.27159} & {-05.27159} & {07.46734} & {-07.46734} & {09.87459} & {-09.87459}
\tabularnewline
 
{4} & {04.54542} & {-04.54542} & {6.65891} & {-06.65891} & {09.58863} & {-09.58863} & {12.71230} & {-12.71230}
\tabularnewline
 
{5} & {05.26144} & {-05.26144} & {8.04526} & {-08.04526} & {11.70748} & {-11.70748} & {15.54566} & {-15.54566}
\tabularnewline
 
{6} & {05.95444} & {-05.95444} & {9.43425} & {-09.43425} & {13.82584} & {-13.82584} & {18.37700} & {-18.37700}\tabularnewline
 
{7} & {06.63386} & {-06.63386} & {10.82655} & {-10.82655} & {15.94430} & {-15.94430} & {21.20730} & {-21.20730}\tabularnewline
 
{8} & {07.30538} & {-07.30538} & {12.22191} & {-12.22191} & {18.06301} & {-18.06301} & {24.03700} & {-24.03700}\tabularnewline
 
{9} & {07.97257} & {-07.97257} & {13.61993} & {-13.61993} & {20.18202} & {-20.18202} & {26.86632} & {-26.86632}\tabularnewline
 
{10} & {08.63771} & {-08.63771} & {15.02016} & {-15.02016} & {22.30130} & {-22.30130} & {29.69541} & {-29.69541}\tabularnewline
 
{11} & {09.30230} & {-09.30230} & {16.42223} & {-16.42223} & {24.42081} & {-24.42081} & {32.52434} & {-32.52434}\tabularnewline
 
{12} & {09.96728} & {-09.96728} & {17.82582} & {-17.82582} & {26.54053} & {-26.54053} & {35.35315} & {-35.35315}\tabularnewline
 
{13} & {10.63327} & {-10.63327} & {19.23069} & {-19.23069} & {28.66043} & {-28.66043} & {38.18189} & {-38.18189}\tabularnewline
 
{14} & {11.30066} & {-11.30066} & {20.63663} & {-20.63663} & {30.78048} & {-30.78048} & {41.01056} & {-41.01056}\tabularnewline
 
{15} & {11.96966} & {-11.96966} & {22.04347} & {-22.04347} & {32.90066} & {-32.90066} & {43.83919} & {-43.83919}\tabularnewline
 
{16} & {12.64037} & {-12.64037} & {23.45108} & {-23.45108} & {35.02095} & {-35.02095} & {46.66778} & {-46.66778}\tabularnewline
 
{17} & {13.31285} & {-13.31285} & {24.85934} & {-24.85934} & {37.14134} & {-37.14134} & {49.49634} & {-49.49634}\tabularnewline
 
{18} & {13.98707} & {-13.98707} & {26.26817} & {-26.26817} & {39.26181} & {-39.26181} & {52.32489} & {-52.32489}\tabularnewline
 
{19} & {14.66300} & {-14.66300} & {27.67749} & {-27.67749} & {41.38235} & {-41.38235} & {55.15342} & {-55.15342}\tabularnewline
 
{20} & {15.34057} & {-15.34057} & {29.08724} & {-29.08724} & {43.50296} & {-43.50296} & {57.98193} & {-57.98193}\tabularnewline
 
{21} & {16.01971} & {-16.01971} & {30.49737} & {-30.49737} & {45.62363} & {-45.62363} & {60.81043} & {-60.81043}\tabularnewline
\hline 
\hline
\end{tabular}
\end{table}
\begin{center}
\begin{figure}[h]
\centering{}\includegraphics[scale=0.6]{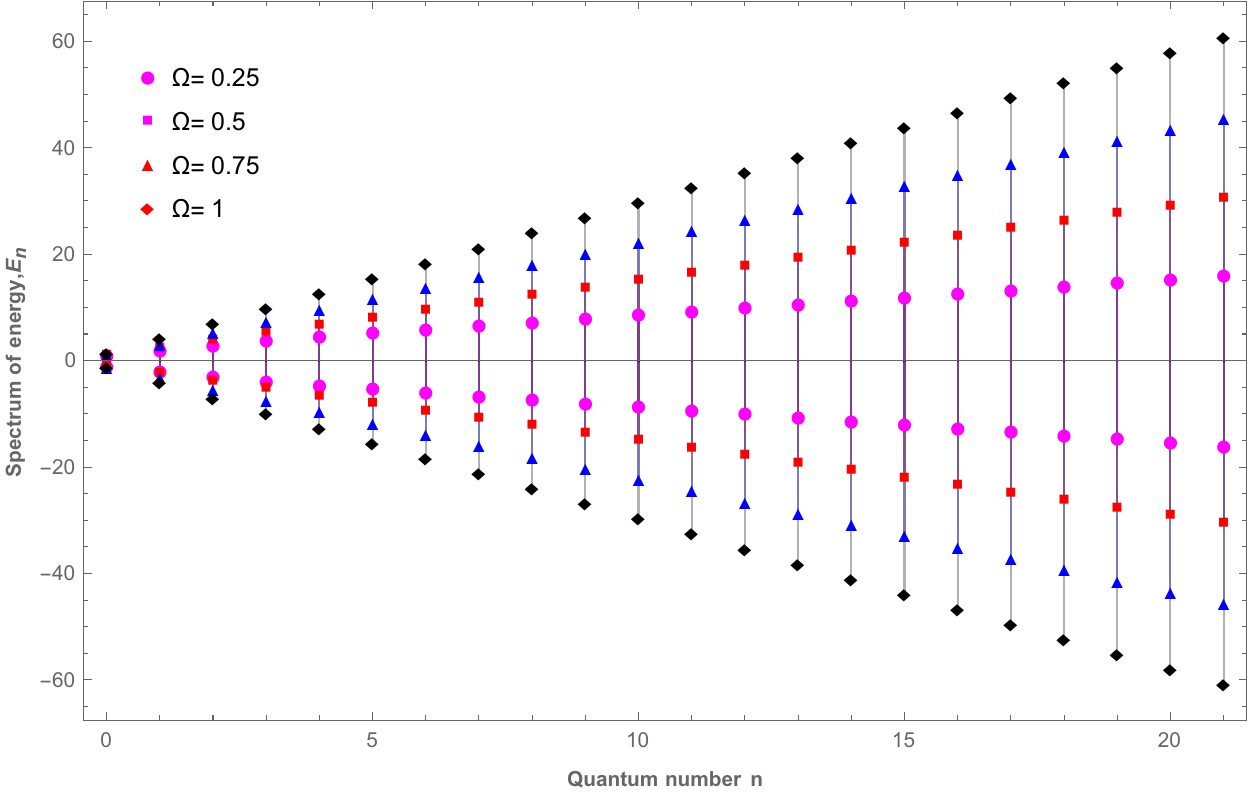}\caption{Plots of the energy of KG oscillator for different values of $\Omega$
with the parameters $K=0$ and $j=m=\omega=1,\varOmega\protect\neq0\, \ell=0.$}
\label{fig.4}
\end{figure}
\par\end{center}
\begin{table}[htpb]
\caption{Some values  of energies with $\ell=1,m=\omega=1,\alpha=0.5,K=0$}
\centering
\begin{tabular}{ccccccccc}
\hline 
\hline 
\textbf{Quantum number, $n$} & \multicolumn{2}{|c|}{{$\Omega=0.25$}} & \multicolumn{2}{c|}{{$\Omega=0.5$}} & \multicolumn{2}{c|}{{$\Omega=0.75$}} & \multicolumn{2}{c|}{{$\Omega=1$}}
\tabularnewline
\hline
\hline 
\multicolumn{1}{c}{
\begin{turn}{} \end{turn}} & {$E^{+}$} & {$E^{-}$} & {$E^{+}$} & {$E^{-}$} & {$E^{+}$} & {$E^{-}$} & {$E^{+}$} & {$E^{-}$}
\tabularnewline
{0} & {02.48636} & {-02.17462} & {03.45262} & {-02.09056} & {05.78575} & {-02.00759} & {09.36895} & {-01.93775}\tabularnewline

{1} & {03.25945} & {-02.94186} & {04.80860} & {-03.22572} & {07.78989} & {-03.61904} & {11.80604} & {-04.01022}\tabularnewline
 
{2} & {04.02474} & {-03.69798} & {06.20194} & {-04.48998} & {09.86040} & {-05.54130} & {14.44309} & {-06.54486}\tabularnewline
 
{3} & {04.75609} & {-04.41863} & {07.59753} & {-05.80549} & {11.95122} & {-07.56557} & {17.15929} & {-09.22120}\tabularnewline
 
{4} & {05.46202} & {-05.11344} & {08.99388} & {-07.14966} & {14.05140} & {-09.63039} & {19.91406} & {-11.95608}\tabularnewline

{5} & {06.15127} & {-05.79174} & {10.39161} & {-08.51195} & {16.15692} & {-11.71484} & {22.69042} & {-14.72094}\tabularnewline
 
{6} & {06.83007} & {-06.46003} & {11.79098} & {-09.88638} & {18.26588} & {-13.81021} & {25.48007} & {-17.50331}\tabularnewline
 
{7} & {07.50259} & {-07.12265} & {13.19200} & {-11.26928} & {20.37723} & {-15.91227} & {28.27850} & {-20.29680}\tabularnewline
 
{8} & {08.17165} & {-07.78247} & {14.59451} & {-12.65828} & {22.49032} & {-18.01872} & {31.08301} & {-23.09781}\tabularnewline
 
{9} & {08.83909} & {-08.44136} & {15.99835} & {-14.05181} & {24.60472} & {-20.12821} & {33.89193} & {-25.90415}\tabularnewline
 
{10} & {09.50617} & {-09.10056} & {17.40334} & {-15.44876} & {26.72014} & {-22.23989} & {36.70413} & {-28.71439}\tabularnewline
 
{11} & {10.17367} & {-09.76085} & {18.80933} & {-16.84838} & {28.83637} & {-24.35321} & {39.51884} & {-31.52759}\tabularnewline
 
{12} & {10.84216} & {-10.42274} & {20.21618} & {-18.25009} & {30.95324} & {-26.46778} & {42.33554} & {-34.34308}\tabularnewline
 
{13} & {11.51193} & {-11.08650} & {21.62377} & {-19.65348} & {33.07065} & {-28.58332} & {45.15380} & {-37.16038}\tabularnewline
 
{14} & {12.18320} & {-11.75228} & {23.03200} & {-21.05824} & {35.18851} & {-30.69965} & {47.97335} & {-39.97913}\tabularnewline
 
{15} & {12.85605} & {-12.42015} & {24.44078} & {-22.46412} & {37.30673} & {-32.81660} & {50.79394} & {-42.79907}\tabularnewline
 
{16} & {13.53053} & {-13.09007} & {25.85006} & {-23.87095} & {39.42526} & {-34.93408} & {53.61541} & {-45.61998}\tabularnewline
 
{17} & {14.20662} & {-13.76202} & {27.25976} & {-25.27856} & {41.54407} & {-37.05199} & {56.43762} & {-48.44172}\tabularnewline
 
{18} & {14.88428} & {-14.43590} & {28.66983} & {-26.68685} & {43.66311} & {-39.17025} & {59.26045} & {-51.26415}\tabularnewline
 
{19} & {15.56346} & {-15.11164} & {30.08024} & {-28.09571} & {45.78234} & {-41.28883} & {62.08381} & {-54.08717}\tabularnewline
 
{20} & {16.24410} & {-15.78913} & {31.49094} & {-29.50506} & {47.90176} & {-43.40767} & {64.90764} & {-56.91070}\tabularnewline
 
{21} & {16.92612} & {-16.46828} & {32.90191} & {-30.91485} & {50.02133} & {-45.52674} & {67.73186} & {-59.73467}\tabularnewline
\hline 
\hline 
\end{tabular}
\end{table}
\begin{center}
\begin{figure}[h]
\centering{}\includegraphics[scale=0.6]{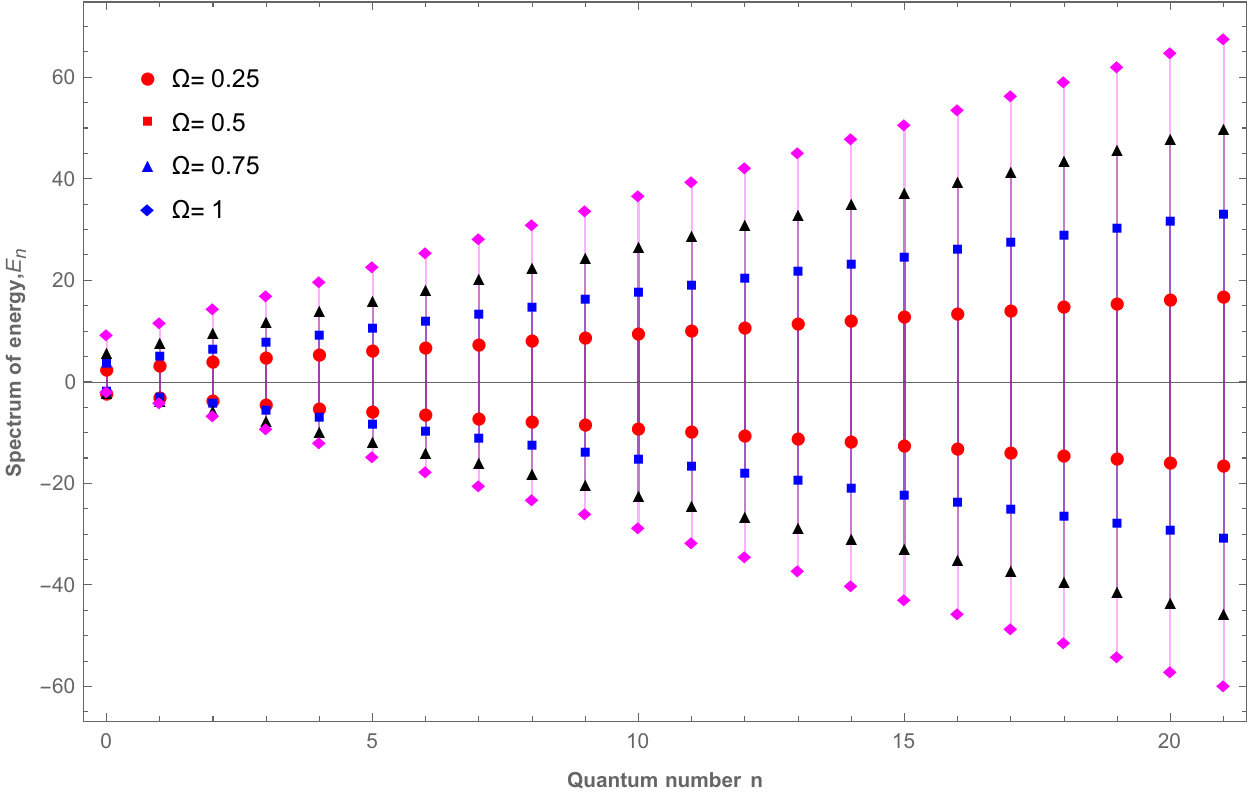}\caption{Plots of the energy of KG oscillator for different values of $\Omega$
with the parameters $K=0$ and $j=m=\omega=1,\varOmega\protect\neq0\;\ell=1.$}
\label{fig.5}
\end{figure}
\par\end{center}
\end{document}